\documentclass[]{article}
\usepackage{graphics}
\usepackage{graphicx}

\begin{document}

\newcommand{\be}{\begin{equation}}
\newcommand{\ee}[1]{\label{#1} \end{equation}}
\newcommand{\ba}{\begin{eqnarray}}
\newcommand{\ea}[1]{\label{#1} \end{eqnarray}}
\newcommand{\nl}{\nonumber \\}
\newcommand{\iint}{\int_0^{\infty}\!}
\newcommand{\pd}[2]{\frac{\partial #1}{\partial #2}}

\renewcommand{\d}[1]{{\rm d}#1}

\title{{\bf Search for light custodians in a clean decay channel at the LHC}}

\author{
 {\sc H.~de Sandes} and {\sc R.~Rosenfeld} \\[1em]
 Instituto de F\'{\i}sica Te\'orica \\ State University of S\~ao Paulo, S\~ao Paulo, Brazil
}

\date{\today}

\maketitle

\begin{abstract}
Models of warped extra dimensions with custodial symmetry 
usually predict the existence of a light Kaluza-Klein fermion
arising as a partner of the right-handed top quark, sometimes called light custodians
which we will denote $\tilde{b}_R$.
The production of these particles at the LHC can give rise to multi-W events
which could be observed in same-sign dilepton channels, but its mass reconstruction 
is challenging.
In this letter we study the possibility of finding
a signal for the pair production of this new particle
at the LHC focusing on a rarer, but cleaner 
decay mode of a light custodian into a $Z$ boson and a $b$-quark. In this mode it would be possible to 
reconstruct the light custodian mass.
 In addition to the dominant standard model QCD
production processes, we include the contribution of a Kaluza-Klein gluon first mode.
We find that the  $\tilde{b}_R$ stands out from the background as a peak in the
$b Z$ invariant mass. However, when taking into account only the electronic and muonic decay 
modes of the $Z$ boson and $b-$tagging efficiencies, 
the LHC will have access only to the very light range of masses, $m_{\tilde{b}} = {\cal O} (500)$ GeV.
\end{abstract}

\section{Introduction}

Models involving an extra space-like 
compact dimension motivated by unification
ideas have a long
history in Physics \cite{KK}. 
More recently, the proposal of a non-factorizable metric
in 5 dimensions, representing a slice of anti-de Sitter AdS$_5$ bulk spacetime
with two 3-branes at its boundaries, 
led to a new possible solution to
the hierarchy problem of the Standard Model (SM) \cite{RS1}.

In the original proposal all SM fields were confined in one
of the branes and only gravity would propagate in the bulk.
Allowing all fields of the SM (with the exception of the Higgs field) to propagate
in the bulk opens up the exciting possibility of explaining
the fermion mass hierarchy with ${\cal O}(1)$ parameters, 
at the same time solving 
the problem of the existence of higher-dimensional operators
which could induce proton decay and flavour violation \cite{GP,GN}.
These models have a very rich phenomenology, with the existence of 
Kaluza-Klein (KK) excitations of the SM particles.  

However, contributions from these new KK modes can violate electroweak
precision constraints described by the S and T parameters \cite{EW} and constraints
from $Z \rightarrow b \bar{b}$.
This situation can be ameliorated by the introduction of a custodial
gauge symmetry in the bulk \cite{custodial1,custodial2}. 
This custodial symmetry has important phenomenological consequences:
it introduces isospin partners of the right handed fermionic fields, sometimes
called custodians \cite{custodians}.
The custodians in some models can have masses not much larger than $500$ GeV.
In particular, these custodians can be produced at the LHC with 
spectacular signatures, such as multi-W events \cite{multiW}.
Finding custodians at the LHC using same-sign dileptons final states coming from multi-W
events was recently studied in \cite{ContinoServant}.

In this letter we describe another signature of the $\tilde{b}_R$ custodian, a charge $-1/3$ 
isospin partner of the right-handed top quark, which is rarer but cleaner than the multi-W signature.

\section{Model and relevant masses and couplings}

There are many different models which predict the existence 
of light custodians. Below we exemplify our strategy with 
a specific minimal model, where the custodians may not 
be particularly light due to their contributions to the $Zbb$ vertex.
However, more complex models can be built in which
this vertex can be protected \cite{BouchartMoreau}.

We will follow the notation of \cite{AgasheServant} and adopt the 5D metric
\begin{equation}
ds^2 =\left( \frac{z_h}{z}\right)^2 \left[ \eta_{\mu \nu} dx^\mu dx^\nu + dz^2 \right]
\end{equation}
where $z$ is the compact 5th dimension bounded by $z_h \leq z \leq z_v$, where 
$z_h = 1/k \approx M_{Pl}^{-1}$ and $z_v = e^{k \pi R}/k \approx \mbox{TeV}^{-1}$ 
are the locations of the Planck and TeV branes,
respectively. For definiteness we will use $k R = 10$.

The model with custodial symmetry has a bulk $SU(2)_L \times SU(2)_R$ and
the 3rd generation quarks are \cite{custodial1}:
\begin{equation}
Q_L = 
\left( \begin{array}{c}
 t_L\\ 
b_L
\end{array} \right); \;\;
Q_{R_1} = 
\left( \begin{array}{c}
 t_R\\ 
\tilde{b}_R
\end{array} \right); \;\; 
Q_{R_2} = 
\left( \begin{array}{c}
 \tilde{t}_R\\ 
b_R
\end{array} \right).
\end{equation} 

Here we will be interested in the $\tilde{b}_R$, the $SU(2)_R$ partner of the right-handed top quark.
Using the notation of \cite{AgasheServant}, the 5D $t_R$ is a $(+,+)$ fermion in order to have a zero mode,
which will be identified with the 4D right-handed top quark. Requiring that the  $SU(2)_L \times SU(2)_R$ is
broken by boundary conditions only at the Planck brane determines $\tilde{b}_R$ to be a $(-+)$ fermion.

The KK decomposition of a 5D fermion is:
 \begin{equation}
\Psi(x,z)=\sum_{n=0}^{\infty }[\psi^{(n)}(x)\chi_n(c,z)],
\end{equation}
where the parameter $c$, related to the fermion bulk mass, determines the profile
of the fermion in the extra dimension. For instance, the profile for the zero-mode fermion
is given by:
\begin{equation}
\chi_0(c,z) = \sqrt{ \frac{1-2 c}{z_h \left(e^{k \pi R (1-2 c)} -1 \right) }} \left(\frac{z}{z_h}\right)^{2-c}.
\end{equation}

The profile for a $(-+)$ fermion is given by \cite{AgasheServant}
\begin{equation}
\chi_n(c,z) = \frac{1}{N_m \sqrt{\pi R}} \left[J_\alpha(m_n z) + b_{\alpha,n} Y_\alpha(m_n z) \right] \left(\frac{z}{z_h}\right)^{5/2},
\end{equation}
where $N_m$ is a normalization factor, $\alpha = |c+1/2|$, and $J_\alpha$ and $Y_\alpha$ are Bessel functions of order $\alpha$.
The mass of a given fermion mode $m_n$ and the constant $b_{\alpha,n}$ are obtained by solving the equations:
\begin{equation}
\frac{J_\alpha(m_n z_h)}{Y_\alpha(m_n z_h)} = \frac{J_{\alpha \mp 1}(m_n z_v)}{Y_{\alpha \mp 1}(m_n z_v)} = -b_{\alpha,n},
\end{equation}
where the upper (lower) signs are for $c > -1/2$ $(c < -1/2)$.

Since the right-handed top must be localized near the TeV brane and $SU(2)_R$ is preserved in the bulk, we will
use $c_{\tilde{b}_R} = c_{t_R} = -1/2$ in the following. With these assignments we find $m_{\tilde{b}_R} = 0.255 \; z_v^{-1}$
and $b_{0,1} = 0.0477$. 

In this letter we will be interested in the pair production of $\tilde{b}_R$. The dominant processes involves QCD interactions.
Due to the fact that the gluon zero-mode profile is flat in the extra dimension, the coupling of the color-triplet  $\tilde{b}_R$ to
gluons is exactly the standard model one. However, one should in principle also compute the couplings to $g^{(1)}$, the first gluonic
KK excitation. Its mass is found to be $m_{g^{(1)}} = 2.40 \; z_v^{-1}$ and the coupling is given by
\begin{equation}
g_{\tilde{b} \tilde{b} g^{(1)}} = g_{QCD} \sqrt{\pi R} \int_{z_h}^{z_v} dz \; \left(\frac{z_h}{z}\right)^4 \chi_1^2(c,z) f_1(z),
\end{equation} 
where $f_1(z)$ is the first gluonic KK excitation profile (see \cite{AgasheServant}). The integrand is peaked towards the TeV brane
and we find a large coupling, $g_{\tilde{b} \tilde{b} g^{(1)}} = 5.45 \; g_{QCD}$. 
This large coupling implies that the KK gluon is a broad resonance, with a width of the order of half of its mass.
For light quarks, however, we find that the coupling $g_{q q g^{(1)}} = 0.2 \; g_{QCD}$
since they are peaked at the Planck brane.

The $\tilde{b}_R$ main decay modes ($\tilde{b}_R \rightarrow b Z, t W, b H$) 
can be studied from the Yukawa interactions of fermions with the bi-doublet Higgs
localized in the TeV brane:
\begin{equation}
S_{Yuk} = \int d^4x \; dz \; \delta(z-z_v) \; H  \left[ \lambda_{U5} \bar{Q}_L Q_{R_1} + 
\lambda_{D5} \bar{Q}_L Q_{R_2} + h.c. \right].
\end{equation}
After making the usual Higgs field re-scaling and identifying the 4D 
top quark Yukawa coupling $\lambda_t$,
we find that there is a $H \tilde{b}_R b_L$ resulting in the
non-diagonal mass matrix:
\begin{equation}
M =  \left( \begin{array}{cc}
 m_{\tilde{b}} & m_t  \\ 
0 & m_b
\end{array}   \right).
\end{equation}
This is very important since it induces a mixing between $\tilde{b}_R$ and $b_L$. It is through this mixing
that the decay modes $\tilde{b}_R \rightarrow b_L Z, t_L W$ are generated with a strength 
(the  $\tilde{b}_R \rightarrow t_R W$ decay could also
occur via a mixing $W-W_R$ but it is suppressed by the small parameter $m_W/m_{W_R}$)  :
\begin{equation}
g_{\tilde{b}_R  b Z} =  \frac{m_t}{m_{\tilde{b}}} g_{bbZ} ; \;\;
g_{\tilde{b}_R  t W} =  \frac{m_t}{m_{\tilde{b}}} g_{btW}.
\end{equation}
On the other hand, the decay $\tilde{b}_R \rightarrow b H$ is generated with a strength set by the top 
Yukawa coupling $\lambda_t \approx 1$.
It turns out that the dominant decay mode is $\tilde{b}_R \rightarrow t_L W$, which generates multi-W
events in the final state and was analyzed in \cite{multiW}.   Here we will concentrate on the study
of the rarer, but cleaner decay $\tilde{b}_R \rightarrow b_L Z$.
We now have all the ingredients necessary to implement and study the $\tilde{b}_R$ pair production at the LHC
in this particular decay channel: $p p \rightarrow \tilde{b}_R \bar{\tilde{b}}_R \rightarrow b Z \bar{b} Z$ .

\section{Model implementation and results}

We implemented the new particles ($\tilde{b}_R$ and $g^{(1)}$) and couplings
in MadGraph/MadEvent \cite{madgraph}.
Fixing $kR=10$, the only free parameter left can be taken as the mass $m_{\tilde{b}}$ of  $\tilde{b}_R$.
We will study the case for $m_{\tilde{b}} = 500, 700$ and $900$ GeV.
We use the total widths calculated in \cite{multiW}.

One should be careful that these mass values do not lead to violations
in the well measured $Zbb$ couplings. In fact, one generally finds
that the contribution arising from fermionic KK partners is given by \cite{custodial2}
\begin{equation}
\delta g_{Zbb} = \sin^2 \theta_{KK} \left(T_L^{3(KK)} - T_L^3 \right)
\end{equation}
where typically the mixing angle is given by
\begin{equation}
 \sin \theta_{KK} \approx \frac{m_t}{m_{KK}}
\end{equation}
which would naivelly require $m_{KK} \geq 1$ TeV. 
However, in more realistic models for the custodians
their masses could be as low as $500$ GeV \cite{custodians}.
 
We included the KK gluon contribution and found it to be  
small due to the fact that it only arises in $s-$channel diagrams initiated by
light quarks, which is suppressed at the LHC and also because the KK gluon is a very broad resonance.

The dominant contribution arises from $t-$channel $\tilde{b}_R$ exchange in gluon-initiated processes.
We have generated the irreducible SM background again using MadGraph/MadEvent.
 
\begin{figure}[h,t,b]
\includegraphics[width=1\textwidth,angle=0]{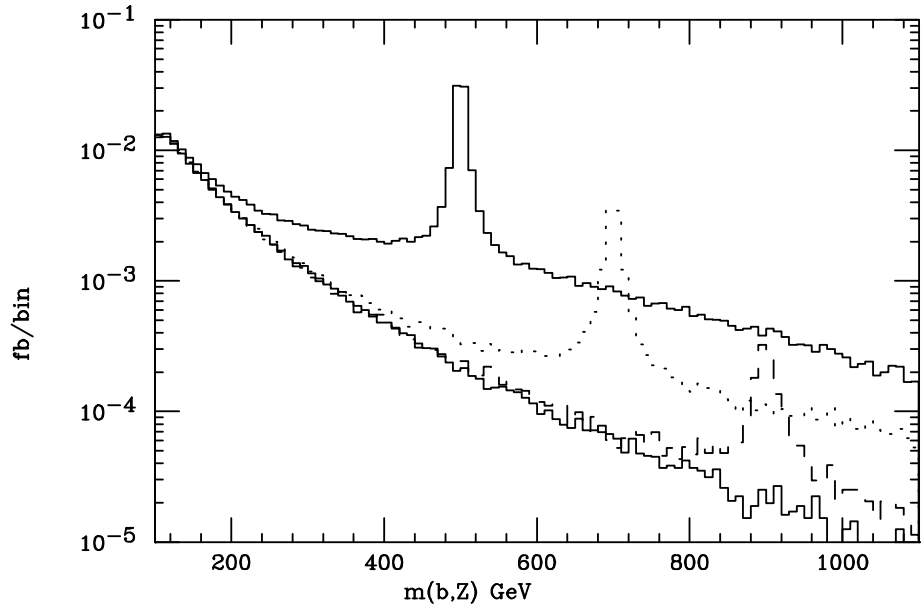}
\caption{\label{1}
Cross section per $10$ GeV bin of the invariant $bZ$ pair mass for the process 
$p p \rightarrow \tilde{b}_R \bar{\tilde{b}}_R \rightarrow b Z \bar{b} Z$
at the LHC for $m_{\tilde{b}} = 500, 700$ and $900$ GeV. Solid line is the SM irreducible background.
}

\end{figure}

In Figure \ref{1} we show the cross section per $10$ GeV bins as a function of the invariant mass of the $b Z$ pair.
We have included the 4 possible combinations of constructing the $b Z$ pair in our analysis.
We have also conservatively assumed that only the ``golden modes" 
 $Z \rightarrow e^+ e^-, \mu^+ \mu^-$ could be used
to reconstruct the $Z$ boson. This implies in a reduction factor of 
$\left( \mbox{BR}(Z \rightarrow  e^+ e^-, \mu^+ \mu^-) \right)^2 = 4.36 \times 10^{-3}$.
We adopted a $b-$tagging efficiency of $50 \%$.  Only a mild $p_T^b > 5$ GeV and $|\eta|< 2.5$ cut was employed.
The $\tilde{b}_R$ signal appears clearly as peaks in the $b Z$ invariant mass but
unfortunately the cross section is small, due mostly to the reduction factors
applied here.

We calculate the luminosity necessary to obtain a $5\sigma$ signal
by integrating the cross section
in a $\pm 100$ GeV window around the resonance peak and
using the estimator $S_{cL}$ proposed in \cite{cms}. It is appropriate for 
a Poisson distribution and hence can be used in processes with a
small number of events. The general expression for the luminosity
necessary to obtain a $N \sigma$ signal is given by
\begin{equation}
    {\cal L}_{N \sigma} = \frac{N^{2}}{2}[ \sigma_{S+B}
    \ln(\frac{ \sigma_{S+B} }{ \sigma_{B}}) -
    (\sigma_{S+B}-\sigma_{B}) ]^{-1}
\end{equation}
where $\sigma_{S+B}$ is the signal plus background cross section
integrated in the mass window and $\sigma_{B}$ is the cross section
only for the background. For the case of a $m_{\tilde{b}_{R}}=500$
GeV we find ${\cal L}_{5 \sigma} = 52.9$ fb$^{-1}$. This luminosity will
produce $5$ signal events over no background. It is possible to lower
this luminosity applying cuts on the b-quark $p_{T}$ e.g. for a $30$
GeV and $|\eta|< 2.5$ cut the luminosity becomes $46.7$ with $3$ signal events.
For $m_{\tilde{b}_{R}}=700$ and $900$ GeV we found luminosities 
greater than $500$ $fb^{-1}$ being
impossible in these cases to observe it on the LHC first years.
However, we should note that these luminosities can drop
by a factor of $20$ if $Z$ decays into $\tau$'s and
neutrinos could be efficiently used.

\begin{figure}[h,t,b]
\includegraphics[width=1\textwidth,angle=0]{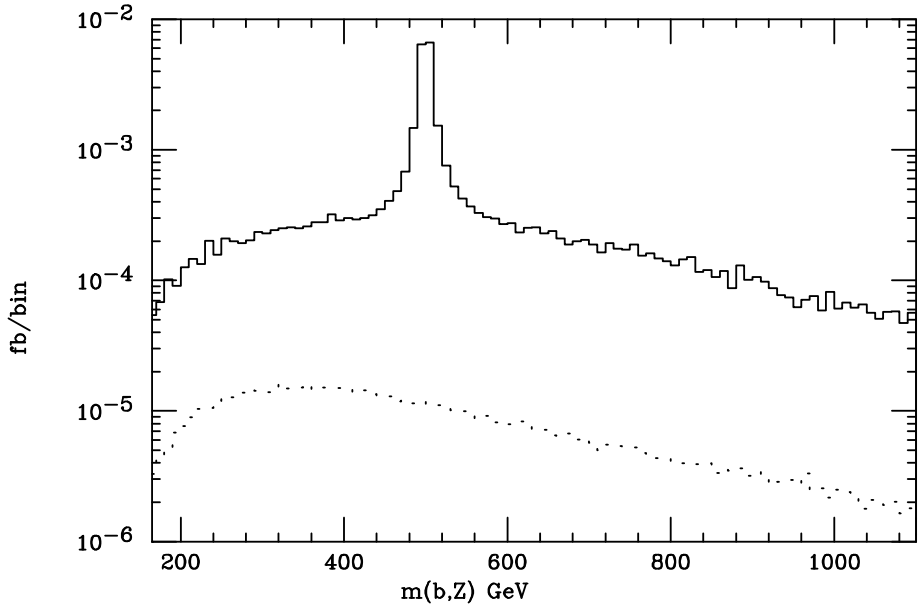}
\caption{\label{2}
Cross section per $10$ GeV bin of the invariant $bZ$ pair mass for the process 
$p p \rightarrow \tilde{b}_R \bar{\tilde{b}}_R \rightarrow b Z \bar{b} Z$
at the LHC for $m_{\tilde{b}} = 500$ GeV with a cut $p_T^b > m_{\tilde{b}}/2$. 
Dotted line is the SM irreducible background.
}

\end{figure}

We also studied the effect of a harder cut  $p_T^b > m_{\tilde{b}}/2$.
In this case, the SM background is dramatically reduced as exemplified in
Figure \ref{2}.
However, in this case the signal cross section around the mass window will
also decrease and even higher luminosities will be necessary to produce few
events.  For $m_{\tilde{b}_{R}}=500$ GeV we find $3$ signal events 
for a $154$ fb$^{-1}$ luminosity.

\section{Conclusions}

We studied the pair production at the LHC of a heavier bottom-like
quark $\tilde{b}_{R}$ which are generically predicted in models of
a warped extra dimension with custodial symmetry in the bulk. We analyzed
the possibility of detection of this new particle focusing on its decay into $Z$ boson and
the b-quark. This decay is suppressed by the  $\tilde{b}_{R}$ mass, which is allowed
to be as low as a few hundreds GeV without inducing large
contributions to $Z\rightarrow\bar{b}b$. This protection is
guaranteed by the custodial symmetry that requires this particle to be in
a multiplet with the right-handed top quark.

We concentrated in the "golden
modes" $Z\rightarrow e^{+}e^{-},\mu^{+}\mu^{-}$ that would allow in
principle a straightforward particle mass reconstruction. We
calculated the differential cross section for the process $pp\rightarrow
\bar{\tilde{b}_{R}}\tilde{b}_{R}\rightarrow Z\bar{b}Zb$ as a
function of the $bZ$ invariant mass taking into account the $Z$
"golden modes" decay branching ratios and b-tagging efficiency. 
We selected three values of the $\tilde{b}_{R}$  mass, namely
$m_{\tilde{b}}=500,700$ and $900$ GeV and different cuts on the
b-quark $p_{T}$. We also took into account the Standard Model
irreducible background. The $\tilde{b}_{R}$ signal appears as a clear peak
in the $Zb$ invariant mass and the luminosity necessary to obtain a
$5\sigma$ signal around this peak was calculated. For the
$m_{\tilde{b}}=500$ GeV case with acceptance cuts of $p_{T}^{b}>5$  and
$30$ GeV and $|\eta|<2.5$, the required luminosities can be
achieved at the LHC first years, but unfortunately for
$m_{\tilde{b}}=700$ and $900$ GeV much higher luminosities are needed.
We also applied a harder cut $p_{T}^{b}>m_{\tilde{b}}/2$ but even
with a significat reduction of the Standard Model background we found that the
signal cross section also decreases and very high luminosities are
necessary to obtain few events. However, further studies with more
realistic simulations must be done, in particular including other $Z$ decay modes 
such as decays into $\tau$'s, neutrinos and even jets (which will have
a much larger background), 
in order to conclude whether this
is an experimental viable decay channel the study of this light
custodian.

We should also emphasize again that in this study we adopted a particularly simple minimal model
of warped extra dimensions with custodial symmetry.
However, more involved models are also possible \cite{BouchartMoreau}.
Unfortunately, the branching ratio of the decay process $\tilde{b}_{R} \rightarrow b Z$
is very model dependent. Hence, a more model-independent analysis, such as
recent work looking for heavy vector-like quarks at the Tevatron \cite{Atre},
should be conducted. Work along these lines is in progress.

\section*{Acknowledgments}
The work of H.~de Sandes is funded by a FAPESP doctoral fellowship.
R.~Rosenfeld thanks CNPq for partial financial support.
We thank  Johan Alwall, Rikkert Frederix, Michel Herquet and Alfonso Zerwekh 
for help in questions related to MadGraph.
We thank Leandro Da Rold and specially G\'{e}raldine Servant for raising important
comments.

\end{document}